\pdfoutput=1

\documentclass[11pt]{article}

\usepackage[final]{acl}
\usepackage{times}
\usepackage{latexsym}
\usepackage{amssymb}
\usepackage{pifont}
\usepackage{booktabs} 
\usepackage{threeparttable} 
\usepackage{graphicx} 
\usepackage{setspace}
\newcommand{\cmark}{\ding{51}} 
\newcommand{\xmark}{\ding{55}} 
\usepackage[T1]{fontenc}

\usepackage[utf8]{inputenc}
\usepackage{multirow}
\usepackage[table,xcdraw]{xcolor}
\usepackage{microtype}

\usepackage{inconsolata}

\usepackage{graphicx}

\definecolor{nightblue}{RGB}{25, 25, 112} 
\definecolor{orangered}{RGB}{255, 69, 0}  

\usepackage{amsmath,amsfonts}
\usepackage{algorithmic}
\usepackage{algorithm}

\title{UniSonate: A Unified Model for Speech, Music, and Sound Effect Generation with Text Instructions}

\author{
 \textbf{Chunyu Qiang\textsuperscript{1,2}},
 \textbf{Xiaopeng Wang\textsuperscript{2}},
 \textbf{Kang Yin\textsuperscript{2}},
 \textbf{Yuzhe Liang\textsuperscript{2}},
 \textbf{Yuxin Guo\textsuperscript{2,3}},
\textbf{Teng Ma\textsuperscript{2}}, 
 \textbf{Ziyu Zhang\textsuperscript{2}},\\
 \textbf{Tianrui Wang\textsuperscript{1}},
 \textbf{Cheng Gong\textsuperscript{1}},
 \textbf{Yushen Chen\textsuperscript{2}},
 \textbf{Ruibo Fu\textsuperscript{3}},
 \textbf{Chen Zhang\textsuperscript{2}},
 \textbf{Longbiao Wang\textsuperscript{1*}},
 \textbf{Jianwu Dang\textsuperscript{1}}
\\
 \textsuperscript{1}Tianjin University, Tianjin, China\\
 \textsuperscript{2}Kling Team, Kuaishou Technology, Beijing, China\\
 \textsuperscript{3}Institute of Automation, Chinese Academy of Sciences, Beijing, China
\\
 \small{
   \{\href{qiangchunyu@tju.edu.cn}{qiangchunyu}, \href{longbiao_wang@tju.edu.cn}{longbiao\_wang}\}@tju.edu.cn
 }
}

\begin{document}
\maketitle
\begin{abstract}
Generative audio modeling has largely been fragmented into specialized tasks, text-to-speech (TTS), text-to-music (TTM), and text-to-audio (TTA), each operating under heterogeneous control paradigms. Unifying these modalities remains a fundamental challenge due to the intrinsic dissonance between structured semantic representations (speech/music) and unstructured acoustic textures (sound effects). In this paper, we introduce \textbf{UniSonate}, a unified flow-matching framework capable of synthesizing speech, music, and sound effects through a standardized, reference-free natural language instruction interface. To reconcile structural disparities, we propose a novel dynamic token injection mechanism that projects unstructured environmental sounds into a structured temporal latent space, enabling precise duration control within a phoneme-driven Multimodal Diffusion Transformer (MM-DiT). Coupled with a multi-stage curriculum learning strategy, this approach effectively mitigates cross-modal optimization conflicts. Extensive experiments demonstrate that UniSonate achieves state-of-the-art performance in instruction-based TTS (WER 1.47\%) and TTM (SongEval Coherence 3.18), while maintaining competitive fidelity in TTA. Crucially, we observe \textit{positive transfer}, where joint training on diverse audio data significantly enhances structural coherence and prosodic expressiveness compared to single-task baselines. Audio samples are available at \url{https://qiangchunyu.github.io/UniSonate/}.
\end{abstract}

\renewcommand{\thefootnote}{\fnsymbol{footnote}} 
\footnotetext{$*$ Corresponding author.} 
\footnotetext{The name ``Sonate'' is derived from the musical term ``Sonata'', symbolizing the model's comprehensive capabilities in audio generation.} 
\begin{figure}[ht]
  \centering
  \includegraphics[width=\linewidth]{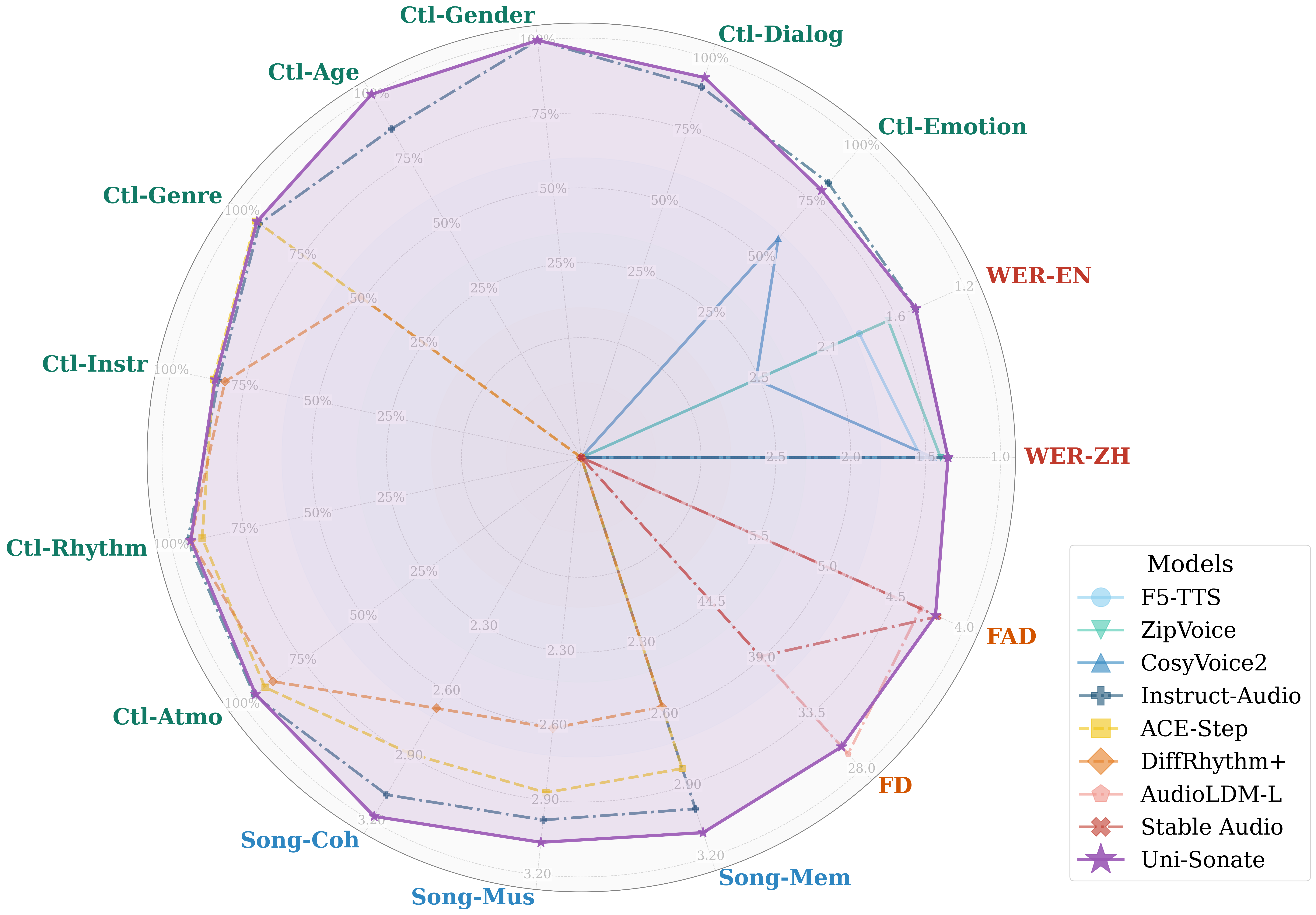}
  \caption{Holistic capability assessment across Speech (SeedTTS-WER, Control), Music (SongEval), and Sound Effects (FAD, FD). Unlike specialized baselines restricted to specific domains, UniSonate achieves \textit{pan-modal coverage}. It demonstrates superior instruction-following and structural coherence in structured tasks (TTS/TTM) while effectively extending to unstructured TTA.}
  \label{fig:compare}
\end{figure}

\section{Introduction}
\label{sec:introduction}
The landscape of neural audio generation has long been fragmented. While specialized models for Text-to-Speech (TTS) \cite{chen2024f5, du2024cosyvoice}, Text-to-Music (TTM) \cite{copet2023simple, gong2025ace}, and Text-to-Audio (TTA) \cite{liu2023audioldm} have achieved remarkable fidelity, they operate under heterogeneous control paradigms. TTS systems typically demand reference audio for timbre cloning and strict phoneme alignment; TTM models rely on lyrics or specialized tags; whereas TTA models generate unstructured textures from open-ended captions. This fragmentation creates a significant barrier to developing general-purpose audio intelligence capable of synthesizing complex auditory scenes—such as dialogue overlaid with background music and environmental effects—within a single probabilistic framework.

Previous attempts at unification have faced substantial limitations regarding consistency and coverage. Models like Vevo2 \cite{zhang2025vevo2} and CosyVoice \cite{du2024cosyvoice} unify speech and singing but remain dependent on reference audio for timbre control, lacking the flexibility of natural language description. UniAudio \cite{yang2023uniaudio} and AudioBox \cite{vyas2023audiobox} support multiple tasks but resort to inconsistent input formats or task-specific fine-tuning, failing to achieve a truly unified interface. To date, no single framework has simultaneously achieved (1) unified generation of speech, music, and sound effects, (2) a consistent instruction-only input format, and (3) reference-free control over fine-grained acoustic attributes.
Achieving this unification presents a fundamental challenge: the intrinsic dissonance between \textit{structured} and \textit{unstructured} semantic representations. Speech and music require precise temporal alignment between discrete units (phonemes/notes) and acoustic realization. Conversely, sound effects (SFX) are inherently holistic and unstructured, lacking rigid temporal boundaries. Simply training a model on concatenated datasets often leads to negative transfer, where the variance of unstructured sound effects destabilizes the articulation required for high-quality speech. While InstructAudio \cite{qiang2025instructaudio} successfully bridged speech and music via structured instruction control, the integration of unstructured environmental sounds remains an unresolved optimization conflict.

In this paper, we introduce \textbf{UniSonate}, a unified generative framework based on conditional flow matching that synthesizes speech, music, and sound effects through a standardized interface. Unlike previous unified models limited to structured modalities, UniSonate introduces a generic alignment paradigm that effectively leverages unstructured environmental audio to bootstrap the performance of structured speech synthesis (Positive Transfer). To reconcile the structural disparities, we propose a novel \textit{Instruction-Content Alignment} paradigm. Beyond mere format standardization, this paradigm seeks to align the semantic space of natural language instructions (e.g., "raspy male voice, sorrowful tone") with the acoustic manifold of diverse audio modalities. We decouple conditioning into two streams: Instruction for high-level attribute control, and Content for temporal structure.

To bridge the gap between discrete linguistic processing and continuous environmental audio, we introduce dynamic token injection. Theoretically, this mechanism acts as the symbolization of unstructured acoustic events, projecting holistic sound effects into a pseudo-linguistic discrete space. By injecting learnable \texttt{[SFX]} tokens, we enable the transformer to process non-verbal audio with the same discrete symbolic reasoning used for phoneme articulation. This allows the model to infer duration and progression for sound effects using shared attention mechanisms, effectively treating all audio generation as a sequence modeling problem.
To harmonize these diverse modalities, UniSonate employs a dual-stream MM-DiT trained via a multi-stage curriculum learning strategy. By progressively expanding from structured speech to semi-structured music and finally to unstructured effects, we mitigate optimization conflicts and catastrophic forgetting.
Our contributions are summarized as follows:
\begin{itemize}
\item We propose UniSonate, the first flow-matching framework to unify TTS, TTM, and TTA tasks under a consistent, reference-free natural language instruction interface, achieving deep semantic alignment between textual descriptions and acoustic features.
\item We introduce Dynamic Token Injection, a mechanism that symbolically represents unstructured acoustic events, enabling precise duration control for sound effects within a phoneme-driven architecture.
\item Extensive experiments demonstrate that UniSonate achieves state-of-the-art performance in instruction-based TTS (WER 1.47\%) and TTM (SongEval Coherence 3.18). Crucially, we observe \textit{positive transfer}: joint training with diverse audio data significantly enhances the structural coherence and prosodic expressiveness of generated speech compared to single-task baselines.
\end{itemize}

\begin{figure*}[t]
 \centering
 \includegraphics[width=\linewidth]{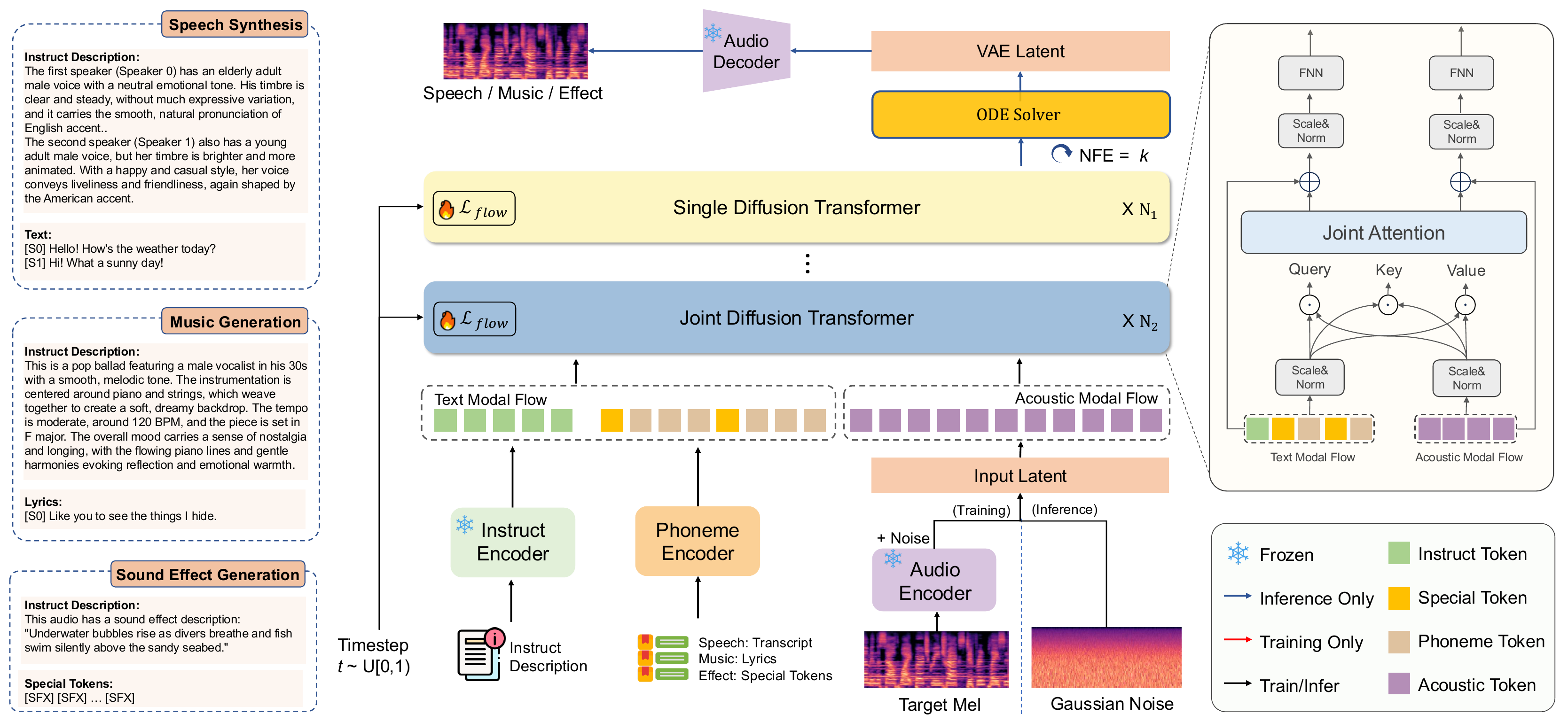}
 \caption{The overall architecture of \textbf{UniSonate}. The framework employs a dual-stream MM-DiT based on conditional flow matching. The input follows the \textit{Instruction-Content Alignment} paradigm, unifying natural language instructions with content sequences, utilizing phonemes for speech/music and special token Injection (via learnable \texttt{[SFX]} tokens) for sound effects. These semantic conditions interact with acoustic latents (compressed by a Mel-VAE) through Joint Diffusion Transformer layers to enable unified audio generation.}
 \label{fig:proposed_model}
\end{figure*}

\section{Related Work}
\subsection{Text-to-Speech}
Driven by generative AI, high-fidelity TTS models based on language modeling (e.g., VALL-E~\cite{wang2023neural, du2024cosyvoice2,qiang2024minimally, cui2025glm}) and flow matching (e.g., F5-TTS~\cite{chen2024f5, wang2025m3, yin2025dmp, qiang2024high}, ZipVoice~\cite{zhu2025zipvoice}) have emerged. While proficient in reference-based cloning, they often lack flexibility. Consequently, research has pivoted to instruction-based control. Pioneers like PromptTTS~\cite{guo2023prompttts} and InstructTTS~\cite{yang2024instructtts} mapped prompts to styles, while ControlSpeech~\cite{ji2024controlspeech} and CosyVoice~\cite{du2024cosyvoice} advanced style-timbre decoupling. Recently, IndexTTS2~\cite{zhou2025indextts2} introduced precise duration mechanisms, and LLM-native frameworks like Spark-TTS~\cite{wang2025spark} and EmoVoice~\cite{yang2025emovoice} leveraged chain-of-thought reasoning for fine-grained prosody control.

\subsection{Text-to-Music}
Text-to-music generation has shifted from symbolic to direct audio synthesis. While early raw-waveform approaches like Jukebox~\cite{dhariwal2020jukebox} proved inefficient, MusicGen~\cite{copet2023simple} established a robust autoregressive framework using discrete tokens, recently scaled by YuE~\cite{yuan2025yue} and SongGen~\cite{liu2025songgen} for full-length songs. Parallelly, latent diffusion models have gained traction for their controllability. AudioLDM 2~\cite{liu2024audioldm} unified audio generation, and MUSTANGO~\cite{melechovsky2024mustango} enhanced attribute control. Notably, the DiffRhythm series~\cite{ning2025diffrhythm,chen2025diffrhythm+} pioneered full-song synthesis with diffusion, achieving fidelity comparable to commercial systems like Suno~\cite{suno2024} and Udio~\cite{udio2024}.

\subsection{Text-to-Audio}
General audio generation has evolved from discrete autoregressive models like AudioGen~\cite{kreuk2022audiogen} to robust latent diffusion approaches exemplified by AudioLDM~\cite{liu2023audioldm} and Make-An-Audio~\cite{huang2023make}. Subsequently, research shifted toward unified foundation models like UniAudio~\cite{yang2023uniaudio} leverages LLM-based tokenization for diverse modalities, while AudioBox~\cite{vyas2023audiobox} employs flow matching to generate speech, music, and sound within a single architecture. Recent works have further expanded cross-modal capabilities: Vevo2~\cite{zhang2025vevo2} bridges speech and singing via unified prosody learning, while Kling-Foley~\cite{wang2025kling} and MMAudio~\cite{cheng2025mmaudio} utilize multimodal Diffusion Transformers to achieve high-fidelity video-to-audio synchronization, demonstrating the potential of complex context modeling.

\section{Method}
\label{sec:method}

Uni-Sonate unifies speech, music, and sound effects within a single probabilistic framework based on conditional flow matching. As shown in Figure \ref{fig:proposed_model}, it employs a dual-stream Multimodal Diffusion Transformer (MM-DiT) that processes a standardized \textit{Instruction-Content} input. The core innovation lies in handling structured (speech/music) and unstructured (SFX) modalities via a unified attention mechanism, enabled by our Dynamic Token Injection strategy.

\subsection{Unified MM-DiT Dual-Stream Architecture}
\label{sec:mmdit}

We employ a MM-DiT architecture underpinned by conditional flow matching \cite{lipman2022flow, qiang2026mm, wang2026apollo}, designed to facilitate bidirectional information flow between semantic conditions and acoustic latents. The architecture is composed of two parallel processing streams—the \textit{Text Stream} and the \textit{Audio Stream}, which interact via joint attention layers.

\noindent\textbf{Text Modality Stream (Conditioning).} 
To unify the heterogeneous control inputs of speech, music, and SFX, we standardize the conditioning signal into a composite sequence. For a given sample, the text stream input $C_{\text{text}}$ is constructed by temporally concatenating the instruction embedding and the content embedding. 
Formally, let $E_I \in \mathbb{R}^{B \times L_I \times D}$ denote the embeddings derived from the natural language instruction (e.g., "A happy male voice," "Upbeat jazz piano," or "Footsteps on gravel"), extracted via a frozen pre-trained instruction encoder (Qwen2.5-7B). 
Let $E_C \in \mathbb{R}^{B \times L_C \times D}$ denote the content embeddings. The nature of $E_C$ varies by task but remains structurally consistent to the transformer.
Speech \& Music: $E_C$ corresponds to the phoneme sequence derived from the transcript or lyrics.
Since SFX lacks linguistic content, $E_C$ is composed of a sequence of learnable \texttt{[SFX]} special tokens. The length of this token sequence is dynamically adjusted to align with the target audio duration, serving as a temporal anchor for the generation process.

The final conditioning input is $C_{\text{text}} = \text{Concat}(E_I, E_C) \in \mathbb{R}^{B \times (L_I+L_C) \times D}$.

\noindent\textbf{Audio Modality Stream (Generation).}
The audio stream processes the noisy latent representations $x_t$. 
Following the SecoustiCodec framework \cite{qiang2025secousticodec, qiang2025vq, qiang2024learning}, we compress 44.1kHz raw waveforms into a compact continuous latent space $x_0$ using a pre-trained Mel-VAE with a downsampling factor of 1024.
During training, $x_t$ represents the linear interpolation between the clean latent $x_0$ and Gaussian noise, following the flow-matching formulation.

\noindent\textbf{Joint Stream Interaction.}
The two streams interact through a stack of $N_2$ Joint Diffusion Transformer layers. In each layer, the text representations $C_{\text{text}}$ and audio latents $x_t$ are processed by separate self-attention blocks to model intra-modal dependencies. Subsequently, a joint attention mechanism concatenates the queries, keys, and values from both modalities, enabling the model to align semantic instructions and content tokens with acoustic textures. This allows the audio stream to attend to the instruction for global style control (e.g., timbre, genre) and the content sequence for fine-grained structural control (e.g., articulation, rhythm).
Following the joint layers, the streams are decoupled. To refine the acoustic details, the audio latents pass through an additional set of $N_1$ Single Diffusion Transformer layers where only self-attention is applied.

\noindent\textbf{Training Objective.}
The model is trained to estimate the vector field $v_\theta$ that transforms the noise distribution to the data distribution. The optimization objective is defined as:
\begin{equation}
    \mathcal{L}_{\text{CFM}} = \mathbb{E}_{t, x_0, x_1, C_{\text{text}}} \big\| v_\theta(t, C_{\text{text}}, x_t) - (x_1 - x_0) \big\|^2
\end{equation}
where $t \in [0, 1]$ is the timestep, and $x_t = t x_1 + (1-t) x_0$. During inference, the target audio latents are reconstructed by integrating the predicted velocity field using an ODE solver (Euler method).

\subsection{Unified Input Representation with Dynamic Special Tokens}
\label{sec:input_rep}

A central challenge in unified audio modeling lies in reconciling the structural disparities between tasks that possess intrinsic linguistic content (speech and music) and those that do not (sound effects). To address this, we propose a standardized input paradigm, the \textit{Instruction-Content Alignment} framework, which extends the instruction-phoneme format of InstructAudio to accommodate the unstructured nature of environmental sounds. As depicted in Figure \ref{fig:proposed_model}, the unified text modality input is composed of two primary segments: a natural language instruction and a content sequence.

The instruction serves as the high-level semantic controller, provided as a natural language prompt. For speech synthesis (TTS), this description specifies speaker attributes such as gender, age, emotion, style, and accent. To support multi-speaker dialogue, we adopt a syntax where distinct descriptions are provided for each speaker, and their respective utterances in the content sequence are prefixed with speaker-id tokens (e.g., \texttt{[S0]}, \texttt{[S1]}). For music generation (TTM), the instruction details musical parameters including genre, instrumentation, tempo, mood, and vocal characteristics (if applicable). For sound effects (SFX), the instruction describes the acoustic event or scene (e.g., "A dog barking in a busy street," "Thunder rolling in the distance").

The content sequence provides fine-grained structural guidance. For TTS and TTM, this is straightforward: the input text or lyrics are converted into a phoneme sequence $C_{\text{text}}$ using a Grapheme-to-Phoneme (G2P) model \cite{qiang2022back}, offering precise temporal alignment for articulation and melody.
However, SFX generation lacks textual transcripts. To integrate SFX into this phoneme-driven architecture without architectural modification, we introduce a dynamic token injection strategy. We define a learnable special token, \texttt{[SFX]}, to serve as a pseudo-phoneme unit. Crucially, the sequence length of these tokens is not arbitrary; it acts as a proxy for temporal duration, enabling the model to infer the length of the audio event.

Let $T_{\text{audio}}$ be the target duration of the sound effect in seconds. We determine the number of special tokens, $L_{\text{sfx}}$, by aligning with the temporal density of speech phonemes. Specifically, we calculate a global scaling factor $\lambda$ from our speech corpus, representing the average phoneme-to-duration ratio:
\begin{equation}
    \lambda = \frac{1}{N} \sum_{i=1}^{N} \frac{\text{len}(P_i)}{\text{duration}(A_i)}
\end{equation}
where $P_i$ and $A_i$ are the phoneme sequence and audio waveform of the $i$-th speech sample, respectively. For any given SFX query with a desired duration $T_{\text{target}}$, the content sequence is constructed as a repetition of the special token:
\begin{equation}
    C_{\text{sfx}} = [\texttt{[SFX]}] \times \lfloor \lambda \cdot T_{\text{target}} \rfloor
\end{equation}

Crucially, we employ repeated tokens rather than a single global \texttt{<duration>} embedding to create temporal anchors. These anchors provide physical "length" in the input space, allowing the MM-DiT's cross-attention to "walk" through the sequence step-by-step. This mechanism mimics the monotonic alignment of phonemes, effectively treating temporal unfolding in SFX as a sequence modeling problem. This design ensures structural integrity for long-form generation and unifies the attention mechanism across all modalities.

\subsection{Multi-Stage Curriculum Learning Strategy}
\label{sec:curriculum}

While the unified architecture enables joint modeling, the intrinsic complexity of the generation tasks varies significantly. Speech synthesis requires high-fidelity capture of linguistic articulation and prosody; music generation demands long-term structural coherence for melody and rhythm; sound effects involve diverse, unstructured acoustic textures. Direct joint training on all modalities from scratch often leads to optimization conflict or negative transfer, where the model struggles to converge on fine-grained speech details due to the high variance of environmental sounds. To mitigate this, we employ a multi-stage curriculum learning strategy (Algorithm \ref{alg:curriculum}). As shown, the training progressively expands from highly structured speech (Stage 1) to semi-structured music (Stage 2), and finally incorporates unstructured sound effects (Stage 3), ensuring robust alignment learning before generalizing to diverse acoustic scenes.

\begin{algorithm}[t]
\caption{Multi-Stage Curriculum Learning Strategy}\label{alg:curriculum}
\begin{algorithmic}
\small
\STATE \textbf{Datasets:} $\mathcal{D}_S$ (Speech), $\mathcal{D}_M$ (Music), $\mathcal{D}_E$ (Effects)
\STATE \textbf{Initialize:} Model parameters $\theta$
\STATE \textbf{Hyperparameters:} $E_1=1$ (Stage 1 Epochs), $E_2=2$ (Stage 2 Epochs)
\STATE $epoch \gets 0$

\WHILE{training not converged}
    \STATE $epoch \gets epoch + 1$
    
    \IF{$epoch \leq E_1$}
        \STATE \textbf{Stage 1: Speech Anchoring}
        \STATE $\mathcal{D}_{curr} \gets \mathcal{D}_S$
    \ELSIF{$epoch \leq E_1 + E_2$}
        \STATE \textbf{Stage 2: Semantic Expansion}
        \STATE $\mathcal{D}_{curr} \gets \mathcal{D}_S \cup \mathcal{D}_M$
    \ELSE
        \STATE \textbf{Stage 3: Universal Generalization}
        \STATE $\mathcal{D}_{curr} \gets \mathcal{D}_S \cup \mathcal{D}_M \cup \mathcal{D}_E$
    \ENDIF
    
    \FOR{batch $B \sim \mathcal{D}_{curr}$}
        \STATE $C_{\text{text}}, x_0 \gets \text{PrepareInput}(B)$
        \STATE Sample $t \sim \mathcal{U}(0, 1)$, $\epsilon \sim \mathcal{N}(0, I)$
        \STATE $x_t \gets t x_1 + (1-t) x_0$
        \STATE $\mathcal{L} \gets \| v_\theta(t, C_{\text{text}}, x_t) - (x_1 - x_0) \|^2$
        \STATE Update $\theta$ via $\nabla_\theta \mathcal{L}$
    \ENDFOR
\ENDWHILE
\end{algorithmic}
\end{algorithm}

\begin{table*}[t]
\centering
\caption{Comprehensive comparison of capabilities across all baselines. UniSonate is the only framework that supports Speech, Music, and Sound Effect generation simultaneously within a single model, while providing the most comprehensive text-based control for speech synthesis.}
\label{tab:capability_comparison}
\resizebox{\linewidth}{!}{
\begin{tabular}{lcccccccccccc}
\toprule
\multirow{2}{*}{\textbf{Model}} & \multirow{2}{*}{\textbf{Params}} & \multirow{2}{*}{\textbf{Data Scale}} & \multicolumn{3}{c}{\textbf{Generation Tasks}} & \multicolumn{6}{c}{\textbf{Control Capabilities}} \\
\cmidrule(lr){4-6} \cmidrule(lr){7-12}
 & & & \textbf{Speech} & \textbf{Music} & \textbf{SFX} & \textbf{Gender} & \textbf{Age} & \textbf{Emo} & \textbf{Style} & \textbf{Accent} & \textbf{Dialogue} \\
\midrule
\multicolumn{12}{l}{\textit{TTS Models}} \\
MaskGCT~\cite{wang2024maskgct} & 1B & 100k hrs (S) & \cmark & \xmark & \xmark & \xmark & \xmark & \xmark & \xmark & \xmark & \xmark \\
E2-TTS~\cite{eskimez2024e2} & 333M & 100k hrs (S) & \cmark & \xmark & \xmark & \xmark & \xmark & \xmark & \xmark & \xmark & \xmark \\
F5-TTS~\cite{chen2024f5} & 336M & 100k hrs (S) & \cmark & \xmark & \xmark & \xmark & \xmark & \xmark & \xmark & \xmark & \xmark \\
ZipVoice~\cite{zhu2025zipvoice} & 123M & 100k hrs (S) & \cmark & \xmark & \xmark & \xmark & \xmark & \xmark & \xmark & \xmark & \xmark \\
CosyVoice1~\cite{du2024cosyvoice} & 416M & 170k hrs (S) & \cmark & \xmark & \xmark & \xmark & \xmark & \cmark & \cmark & \cmark & \xmark \\
CosyVoice2~\cite{du2024cosyvoice2} & 618M & 167k hrs (S) & \cmark & \xmark & \xmark & \xmark & \xmark & \cmark & \cmark & \cmark & \xmark \\
\midrule
\multicolumn{12}{l}{\textit{TTM Models}} \\
DiffRhythm+~\cite{chen2025diffrhythm+} & 1B & 120k hrs (M) & \xmark & \cmark & \xmark & -- & -- & -- & -- & -- & -- \\
ACE-Step~\cite{gong2025ace} & 3B & 100k hrs (M) & \xmark & \cmark & \xmark & -- & -- & -- & -- & -- & -- \\
\midrule
\multicolumn{12}{l}{\textit{TTA Models}} \\
AudioLDM-L~\cite{liu2023audioldm} & 739M & 634k clips (E) & \xmark & \xmark & \cmark & -- & -- & -- & -- & -- & -- \\
Tango-FT~\cite{ghosal2023text} & 866M & 45k clips (E) & \xmark & \xmark & \cmark & -- & -- & -- & -- & -- & -- \\
EzAudio-XL~\cite{hai2024ezaudio} & 875M & 270k clips (E) & \xmark & \xmark & \cmark & -- & -- & -- & -- & -- & -- \\
Stable Audio~\cite{evans2025stable} & 1.0B & 486k clips (E) & \xmark & \xmark & \cmark & -- & -- & -- & -- & -- & -- \\
GenAU-L~\cite{haji2024taming} & 1.2B & 811k clips (E) & \xmark & \xmark & \cmark & -- & -- & -- & -- & -- & -- \\
\midrule
\multicolumn{12}{l}{\textit{Unified Models}} \\
InstructAudio~\cite{qiang2025instructaudio} & 1.3B & \begin{tabular}[c]{@{}c@{}}50k hrs (S)\\ + 20k hrs (M)\end{tabular} & \cmark & \cmark & \xmark & \cmark & \cmark & \cmark & \cmark & \cmark & \cmark \\
\rowcolor{gray!10} \textbf{UniSonate (Ours)} & 1.3B & \begin{tabular}[c]{@{}c@{}}50k hrs (S)\\+ 20k hrs (M)\\+ 1.5M clips (E)\end{tabular} & \cmark & \cmark & \cmark & \cmark & \cmark & \cmark & \cmark & \cmark & \cmark \\
\bottomrule
\end{tabular}
}
\begin{tablenotes}
    \scriptsize
    \item \textbf{Note:} \textbf{S}=Speech, \textbf{M}=Music, \textbf{E}=Sound Effects (clips). \textbf{SFX}: Sound Effects Generation.
\end{tablenotes}
\end{table*}

\begin{table*}[h]
\centering
\caption{Performance comparison of instruction-based TTS on control accuracy, similarity, distortion/error metrics, and subjective evaluation. UniSonate demonstrates superior signal quality and dialogue control while maintaining competitive expressiveness.}
\label{tab:tts_results}
\resizebox{\textwidth}{!}{ 
\begin{tabular}{lccccccccccccccc}
\toprule
\multirow{2}{*}{\textbf{Model}} & \multicolumn{6}{c}{\textbf{Classification Control Accuracy Rate (\%)$\uparrow$}} & \multicolumn{2}{c}{\textbf{Similarity$\uparrow$}} & \multicolumn{4}{c}{\textbf{Distortion/Error $\downarrow$}} & \multicolumn{2}{c}{\textbf{MOS$\uparrow$}} \\
\cmidrule(lr){2-7} \cmidrule(lr){8-9} \cmidrule(lr){10-13} \cmidrule(lr){14-15}
& \textbf{Gender} & \textbf{Age} & \textbf{Emotion} & \textbf{Style} & \textbf{Accent} & \textbf{Dialog} & \textbf{Spk} & \textbf{Emo} & \textbf{LSD} & \textbf{MCD} & \textbf{MSEP} & \textbf{MR} & \textbf{QMOS} & \textbf{NMOS} \\
\midrule
Ground Truth & 100.00 & 100.00 & 100.00 & 100.00 & 100.00 & 100.00 & 1.00 & 1.00 & 0.00 & 0.00 & 0.00 & 0.00 & -- & -- \\
CosyVoice2\cite{du2024cosyvoice2} & -- & -- & 58.33 & 65.00 & \textbf{100.00} & -- & 0.68 & 0.53 & 2.57 & 7.11 & 547.87 & 0.46 & \textbf{3.90 ± 0.11} & \textbf{3.65 ± 0.22} \\
InstructAudio\cite{qiang2025instructaudio} & \textbf{100.00} & \textbf{86.67} & \textbf{83.33} & \textbf{86.67} & \textbf{100.00} & 90.00 & 0.76 & \textbf{0.71} & 1.88 & 5.71 & 437.58 & 0.33 & 3.73 ± 0.24 & 3.46 ± 0.32 \\
\rowcolor{gray!10} \textbf{UniSonate} & \textbf{100.00} & \textbf{86.67} & 80.00 & 80.00 & \textbf{100.00} & \textbf{93.33} & \textbf{0.77} & 0.67 & \textbf{1.79} & \textbf{5.46} & \textbf{422.36} & \textbf{0.31} & 3.83 ± 0.17 & 3.50 ± 0.18 \\
\bottomrule
\end{tabular}
}
\end{table*}

\begin{table}[t]
\centering
\caption{Comparison of Word Error Rate (WER) performance. UniSonate achieves the best recognition accuracy on both English and Chinese datasets.}
\label{tab:wer_results}
\resizebox{0.8\columnwidth}{!}{ 
\begin{tabular}{lcc}
\toprule
\multirow{2}{*}{\textbf{Model}} & \multicolumn{2}{c}{\textbf{WER(\%)$\downarrow$}} \\
\cmidrule(lr){2-3}
& \textbf{EN} & \textbf{ZH} \\
\midrule
Ground Truth & 2.14 & 1.25 \\
MaskGCT\cite{wang2024maskgct} & 2.26 & 2.40 \\
E2-TTS\cite{eskimez2024e2} & 2.49 & 1.91 \\
F5-TTS\cite{chen2024f5} & 1.89 & 1.53 \\
ZipVoice\cite{zhu2025zipvoice} & 1.70 & 1.40 \\
CosyVoice1\cite{du2024cosyvoice} & 4.29 & 3.63 \\
CosyVoice2\cite{du2024cosyvoice2} & 2.57 & 1.45 \\
InstructAudio\cite{qiang2025instructaudio} & 1.52 & 1.35 \\
\midrule
\rowcolor{gray!10} \textbf{UniSonate (Ours)} & \textbf{1.47} & \textbf{1.25} \\
\bottomrule
\end{tabular}
}
\end{table}

\begin{table*}[h]
\centering
\caption{Performance comparison of TTM on control accuracy, SongEval, and subjective evaluation. UniSonate achieves state-of-the-art results on all SongEval metrics and Musicality MOS (MMOS), demonstrating that unified training enhances musical structure.}
\label{tab:music_results}
\resizebox{\textwidth}{!}{ 
\begin{tabular}{lccccccccccccc}
\toprule
\multirow{2}{*}{\textbf{Model}} & \multicolumn{6}{c}{\textbf{Classification Control Accuracy Rate (\%)$\uparrow$}} & \multicolumn{5}{c}{\textbf{SongEval$\uparrow$}} & \multicolumn{2}{c}{\textbf{MOS$\uparrow$}} \\
\cmidrule(lr){2-7} \cmidrule(lr){8-12} \cmidrule(lr){13-14}
& \textbf{Genre} & \textbf{Instr} & \textbf{Gend} & \textbf{Age} & \textbf{Rhy} & \textbf{Atmo} & \textbf{Coh} & \textbf{Mus} & \textbf{Mem} & \textbf{Cla} & \textbf{Nat} & \textbf{QMOS} & \textbf{MMOS} \\
\midrule
Ground Truth & 100.0 & 100.0 & 100.0 & 100.0 & 100.0 & 100.0 & 3.60 & 3.52 & 3.56 & 3.43 & 3.34 & -- & -- \\
DiffRhythm+\cite{chen2025diffrhythm+} & 51.33 & 81.67 & 22.22 & 44.44 & 93.33 & 87.22 & 2.68 & 2.61 & 2.57 & 2.48 & 2.37 & 3.04 ± 0.46 & 2.79 ± 0.54 \\
ACE-Step\cite{gong2025ace} & \textbf{94.44} & \textbf{85.56} & 96.11 & 95.00 & 89.44 & 90.56 & 2.89 & 2.87 & 2.83 & 2.77 & 2.71 & \textbf{3.30 ± 0.28} & 2.88 ± 0.20 \\
InstructAudio\cite{qiang2025instructaudio} & 92.78 & 83.89 & \textbf{98.89} & 97.22 & \textbf{94.44} & \textbf{95.00} & 3.08 & 2.98 & 3.00 & 2.89 & 2.82 & 2.82 ± 0.26 & 2.91 ± 0.35 \\
\midrule
\rowcolor{gray!10} \textbf{UniSonate} & 93.89 & 85.00 & \textbf{98.89} & \textbf{97.78} & 93.33 & 94.44 & \textbf{3.18} & \textbf{3.07} & \textbf{3.10} & \textbf{2.99} & \textbf{2.90} & 2.88 ± 0.21 & \textbf{3.01 ± 0.29} \\
\bottomrule
\end{tabular}
}
\begin{tablenotes}
      \scriptsize
      \item \textbf{Note:} Instr=Instrument, Gend=Gender, Rhy=Rhythm, Atmo=Atmosphere; Coh=Coherence, Mus=Musicality, Mem=Memorability, Cla=Clarity, Nat=Naturalness.
\end{tablenotes}
\end{table*}

\begin{table}[t]
\centering
\caption{Performance comparison on Sound Effects (TTA) generation benchmarks. UniSonate leverages a unified dataset of speech, music, and effects to achieve competitive fidelity.}
\label{tab:tta_metrics}
\resizebox{\linewidth}{!}{
\begin{tabular}{lccccc}
\toprule
\textbf{Model} & \textbf{FAD}$\downarrow$ & \textbf{FD}$\downarrow$ & \textbf{KL}$\downarrow$ & \textbf{IS}$\uparrow$ & \textbf{CLAP}$\uparrow$ \\
\midrule
Ground Truth & 0.00 & 0.00 & 0.00 & -- & -- \\
AudioLDM-L~\cite{liu2023audioldm} & 4.32 & 29.50 & 1.68 & 8.17 & 0.208 \\
Tango-FT~\cite{ghosal2023text} & 2.68 & 15.64 & \textbf{1.24} & 8.78 & 0.291 \\
EzAudio-XL~\cite{hai2024ezaudio} & 3.64 & 14.98 & 1.29 & \textbf{11.38} & \textbf{0.314} \\
Stable Audio~\cite{evans2025stable} & 4.19 & 39.14 & 2.36 & 10.07 & 0.209 \\
GenAU-L~\cite{haji2024taming} & \textbf{2.07} & \textbf{14.58} & 1.36 & 10.43 & 0.300 \\
\midrule
\rowcolor{gray!10} \textbf{UniSonate (Ours)} & 4.21 & 30.21 & 2.44 & 8.22 & 0.156 \\
\bottomrule
\end{tabular}
}
\end{table}

\section{Experiments}

We compare Uni-Sonate against domain-specific SOTA models using standard objective metrics and subjective MOS. Detailed baseline configurations and metric definitions are in Appendix \ref{sec:appendix:setup}.

\subsection{Datasets}
We construct a large-scale unified audio corpus comprising three distinct modalities: speech, music, and sound effects. 
The dataset consists of 50K hours of speech and 20K hours of music collected from internet sources, consistent with InstructAudio, alongside a newly introduced collection of 1.5 million sound effect (SFX) clips.
We apply a standardized internal data processing pipeline to generate unified natural language instructions across all tasks. 
For speech, instructions cover attributes including gender, age, emotion, style, and accent. 
Music instructions detail genre, instrument, rhythm, and atmosphere. 
For the SFX data, instructions describe acoustic events (e.g., "footsteps," "glass breaking") and environmental scenes.
All audio samples are standardized to a 44.1kHz sampling rate, with clip durations ranging from 2 to 20 seconds. The speech data maintains balanced 1:1 ratios for Chinese-English languages and gender distribution, including 0.5\% dialogue-specific data to support multi-speaker generation.

\subsection{Model Architecture}
UniSonate is built upon a MM-DiT architecture comprising approximately 1.34 billion parameters. The model utilizes a flow matching feedforward dimension of 1024 and consists of 14 Joint Diffusion Transformer layers followed by 6 Single Diffusion Transformer layers, incorporating RoPE positional encoding \cite{su2024roformer} for temporal awareness. 
For conditioning, we employ Qwen2.5-7B \cite{qwen2025qwen25technicalreport} as the frozen instruction encoder to process natural language descriptions. The content encoder utilizes a Zipformer-based \cite{zhu2025zipvoice} network (512 dimension) to encode phoneme sequences for speech and music, while employing learnable special tokens for sound effects to model temporal duration.
Audio is processed via a pre-trained Mel-VAE encoder that compresses 44.1kHz waveforms into continuous latent embeddings at 43 Hz, achieving a 1024$\times$ downsampling rate. 
Training is conducted on 32 NVIDIA Tesla A800 80GB GPUs with a batch size of 16 per GPU, utilizing the Adam optimizer \cite{Kingma2014AdamAM} with an initial learning rate of $1e^{-4}$.

\begin{table*}[t]
\centering
\caption{Ablation study on Speech Synthesis (TTS). We compare the full UniSonate model against a variant trained exclusively on speech data with identical architecture. The joint training significantly improves intelligibility (WER) and signal fidelity (LSD/MCD), demonstrating that diverse audio modalities enhance speech robustness.}
\label{tab:ablation_tts}
\resizebox{0.8\linewidth}{!}{
\begin{tabular}{lcccccccc}
\toprule
\textbf{Training Configuration} & \textbf{WER-EN}$\downarrow$ & \textbf{WER-ZH}$\downarrow$ & \textbf{Sim-Spk}$\uparrow$ & \textbf{Sim-Emo}$\uparrow$ & \textbf{LSD}$\downarrow$ & \textbf{MCD}$\downarrow$ & \textbf{MSEP}$\downarrow$ & \textbf{MR}$\downarrow$ \\
\midrule
UniSonate (TTS-Only Data) & 2.24 & 1.40 & 0.63 & 0.51 & 2.63 & 8.70 & 574.67 & 0.426 \\
\rowcolor{gray!10} \textbf{UniSonate (Joint Data)} & \textbf{1.47} & \textbf{1.25} & \textbf{0.77} & \textbf{0.67} & \textbf{1.79} & \textbf{5.46} & \textbf{422.36} & \textbf{0.31} \\
\bottomrule
\end{tabular}
}
\vspace{-0.1in}
\end{table*}

\begin{table}[t]
\centering
\caption{Ablation study on Music Generation (TTM). Comparing the full unified model against a music-only variant. Joint training yields improvements across all SongEval metrics, indicating that large-scale structured speech data helps the model learn better musical coherence.}
\label{tab:ablation_ttm}
\resizebox{\linewidth}{!}{
\begin{tabular}{lccccc}
\toprule
\multirow{2}{*}{\textbf{Training Configuration}} & \multicolumn{5}{c}{\textbf{SongEval$\uparrow$}} \\
\cmidrule(lr){2-6}
& \textbf{Coh} & \textbf{Mus} & \textbf{Mem} & \textbf{Cla} & \textbf{Nat} \\
\midrule
UniSonate (TTM-Only Data) & 3.11 & 3.00 & 3.04 & 2.92 & 2.84 \\
\rowcolor{gray!10} \textbf{UniSonate (Joint Data)} & \textbf{3.18} & \textbf{3.07} & \textbf{3.10} & \textbf{2.99} & \textbf{2.90} \\
\bottomrule
\end{tabular}
}
\vspace{-0.15in}
\end{table}

\subsection{Results and Analysis}
\label{sec:results}

\subsubsection{Evaluation of TTS} 
We first evaluate the fundamental speech generation capabilities. Table \ref{tab:wer_results} reports the Word Error Rate (WER) on the Seed-TTS test set. UniSonate achieves the lowest WER (1.47\% on English and 1.25\% on Chinese), surpassing both the dedicated TTS baselines (e.g., F5-TTS, CosyVoice2) and the previous unified model InstructAudio. This suggests that the inclusion of diverse audio data (music and sound effects) during the curriculum learning phase does not dilute speech intelligibility; rather, it appears to enhance the model's acoustic robustness.

Table \ref{tab:tts_results} presents a detailed comparison of instruction-based control against the SOTA model CosyVoice2 and InstructAudio. UniSonate exhibits superior controllability and signal quality:
UniSonate maintains 100\% accuracy in Gender and Accent control and achieves 93.33\% in Dialogue control—a capability entirely absent in CosyVoice2. Compared to InstructAudio, UniSonate improves dialogue handling.
In terms of distortion metrics, UniSonate achieves the best performance with an LSD of 1.79 and MCD of 5.46. It consistently outperforms CosyVoice2, which suffers from emotion leakage due to its reliance on reference audio.
 While CosyVoice2 achieves a slightly higher QMOS due to reference-based guidance, UniSonate attains a comparable NMOS (3.50) using pure text instructions, significantly reducing the ambiguity inherent in one-to-many mappings.

\subsubsection{Evaluation of TTM} 
Table \ref{tab:music_results} compares UniSonate against specialized music generation models. While specialized models like ACE-Step excel in genre classification accuracy, UniSonate demonstrates superior performance in structural and detailed attributes.
Notably, UniSonate achieves state-of-the-art results on the SongEval benchmark, with the highest scores in Coh (3.18) and Mus (3.07). This represents a significant improvement over InstructAudio (Coh 3.08). We hypothesize that the unified training with large-scale speech data enhances the model's ability to model long-term temporal dependencies, which transfers positively to musical structure. Subjectively, UniSonate achieves the highest Musicality MOS (3.01), validating that our unified architecture captures melodic nuances effectively without specialized music-only architectural designs.

\subsubsection{Evaluation of TTA} 
Table \ref{tab:tta_metrics} assesses the newly added sound effect generation capability. UniSonate achieves an FAD of 4.21 and CLAP score of 0.156, demonstrating competitive performance comparable to widely used baselines such as AudioLDM-L (FAD 4.32) and Stable Audio (FAD 4.19). 
While there is a performance gap compared to the specialized SOTA model GenAU-L, we consider this trade-off acceptable given UniSonate's unique position as a unified multi-task model. Unlike specialized TTA models that focus exclusively on a single modality, UniSonate accommodates speech, music, and sound effects within one framework.
Crucially, the results confirm that UniSonate successfully learns to generate non-linguistic acoustic events using our proposed dynamic token injection strategy. This validates that the multi-stage curriculum learning effectively integrates unstructured sound effects into a phoneme-driven architecture without causing catastrophic forgetting of speech or music capabilities.

Across all three domains, UniSonate demonstrates that a single unified model can achieve performance superior to domain-specific specialists in structured tasks (TTS and TTM) while maintaining competitive fidelity in unstructured tasks (TTA). The improvements over InstructAudio in both TTS and TTM metrics indicate that scaling up data diversity through sound effects and employing curriculum learning leads to positive transfer across varying acoustic modalities, proving the viability of a truly unified audio generation model.

\subsubsection{Effectiveness of Joint Training (Ablation Study)} 
To rigorously validate the superiority of unified modeling over single-task approaches, we conducted a controlled ablation study. We retrained the exact same UniSonate architecture under two restricted data configurations: one using exclusively speech data (TTS-Only) and another using exclusively music data (TTM-Only), while keeping all hyperparameters and model size constant.
As shown in Table \ref{tab:ablation_tts}, the joint-trained UniSonate (Speech+Music+SFX) significantly outperforms its TTS-only counterpart. The English WER drops from 2.24\% to 1.47\%, and spectral fidelity metrics (LSD, MCD) show marked improvements. This confirms that exposing the model to the rich acoustic diversity of music and sound effects enhances its generalization capabilities, allowing the shared encoder to learn more robust acoustic features that benefit speech reconstruction.
Table \ref{tab:ablation_ttm} reveals a similar trend in music generation. The unified model surpasses the TTM-only variant across all SongEval metrics. We attribute this to the inclusion of 50K hours of highly structured speech data. The strict alignment requirements of speech training likely force the model to learn better temporal attention mechanisms, which \textit{positively transfers} to music generation, resulting in improved structural coherence (Coh) and rhythm stability.

\section{Conclusions}

We presented Uni-Sonate, a unified flow-matching framework that synthesizes speech, music, and sound effects under a single architecture. By introducing Dynamic Token Injection and a multi-stage curriculum learning strategy, we successfully harmonized structured and unstructured audio modalities. Our results demonstrate not only state-of-the-art performance in instruction-based TTS and TTM but also, crucially, that unified training induces positive transfer, enhancing the generation quality of individual tasks. Uni-Sonate paves the way for general-purpose audio intelligence capable of complex auditory scene synthesis.

\section{Limitations}
\label{sec:limitations}

While UniSonate demonstrates the potential of a unified framework for speech, music, and sound effect generation, several limitations remain to be addressed in future work.
As shown in Table \ref{tab:tta_metrics}, although UniSonate achieves competitive performance in sound effect generation, there is still a noticeable gap in Fréchet Audio Distance (FAD) compared to specialized SOTA models like GenAU-L (4.21 vs. 2.07). This suggests that while the unified representation is effective, the model may struggle to capture the extreme diversity of unstructured acoustic environments as effectively as models dedicated solely to that modality.
Currently, our training and evaluation focus primarily on audio clips ranging from 2 to 20 seconds. While the model excels at short-context coherence, generating consistent long-form content (e.g., full songs exceeding 3 minutes or extended audiobooks) remains challenging. The attention mechanism's memory constraints and the lack of a hierarchical structure for long-term planning limit the model's ability to maintain musical structure or narrative consistency over extended durations.
Relying solely on natural language instructions introduces inherent one-to-many mapping ambiguity. Unlike reference-based methods that provide explicit acoustic cues, text descriptions (e.g., "a sad song") can correspond to vastly different acoustic realizations. This sometimes results in generations that, while faithful to the text, may not align with the user's specific unstated preferences, leading to slight variances in perceived naturalness compared to reference-conditioned systems.
As a 1.3B parameter diffusion model requiring multiple denoising steps, UniSonate is computationally intensive during inference compared to lightweight, non-autoregressive TTS systems. This currently limits its applicability in real-time scenarios requiring low-latency synthesis.

\section{Ethical Considerations}
\label{sec:ethics}

The development of high-fidelity unified audio generation models brings significant capabilities but also necessitates careful consideration of potential risks and ethical implications.
UniSonate's ability to generate realistic speech and dialogue via text instructions poses a risk of misuse for creating misleading content, disinformation, or "deepfakes." Although our model relies on descriptive prompts (e.g., "young male") rather than direct voice cloning from reference audio—which theoretically reduces the risk of impersonating specific individuals without their consent—the high quality of the output could still be exploited to deceive listeners.
Our model is trained on large-scale datasets collected from the internet. Consequently, it may inherit biases present in the training data, such as gender stereotypes associated with certain professions in speech, or Western-centric biases in musical genres. There is a risk that the model may default to these biases when instructions are underspecified. We are committed to further analyzing these biases and developing methods to ensure more equitable representation.
The music generation capability raises concerns regarding copyright and artistic style mimicry. While the model generates original compositions based on text, the training process utilizes existing musical works. We emphasize that this tool is intended to assist creators rather than replace human artists. Future releases will strictly adhere to copyright laws, and we are exploring mechanisms such as dataset filtering and output watermarking to respect intellectual property rights.
o mitigate these risks, we plan to release the model weights under a license that prohibits malicious use. Furthermore, we advocate for the development and integration of synthetic audio detection tools (watermarking) to help users distinguish between human-produced and AI-generated audio content.

\section{Acknowledgement}
\label{sec:acknowledgement}
This work was supported by the National Natural Science Foundation of China under Grant U23B2053.

\bibliography{custom}

\appendix

\section{Compared Methods and Evaluation Metrics}
\label{sec:appendix:setup}

To evaluate UniSonate's unified generation capabilities across speech, music, and sound effects, we compare against state-of-the-art (SOTA) specialized models in each domain as well as the previous unified model, InstructAudio\cite{qiang2025instructaudio}.

\subsection{Baselines}
For TTS, we benchmark fundamental generation quality against MaskGCT\cite{wang2024maskgct}, E2-TTS\cite{eskimez2024e2}, F5-TTS\cite{chen2024f5}, ZipVoice\cite{zhu2025zipvoice}, CosyVoice1\cite{du2024cosyvoice}, and CosyVoice2\cite{du2024cosyvoice2} (Table \ref{tab:capability_comparison} \& \ref{tab:wer_results}). We specifically compare instruction-based control performance against CosyVoice2 and InstructAudio (Table \ref{tab:tts_results}). 
Consistent with previous settings, since UniSonate is purely instruction-controlled, we use neutral text descriptions with randomized speakers for Seed-TTS WER evaluation. For CosyVoice2, which requires reference audio for timbre, we provide matching reference samples and map instructions to its supported control tags.
For Music (TTM), we compare with DiffRhythm+\cite{chen2025diffrhythm+}, ACE-Step\cite{gong2025ace}, and InstructAudio (Table \ref{tab:music_results}). As DiffRhythm+ lacks support for short-duration synthesis, we generate longer sequences and truncate them for fair comparison.
For Sound Effects (TTA), we benchmark against specialized latent diffusion models including AudioLDM-L\cite{liu2023audioldm}, Tango-FT\cite{ghosal2023text}, EzAudio-XL\cite{hai2024ezaudio}, Stable Audio\cite{evans2025stable}, and GenAU-L\cite{haji2024taming} (Table \ref{tab:tta_metrics}).

\subsection{Evaluation Metrics}
We employ a comprehensive suite of objective and subjective metrics tailored to each modality.
Speech Metrics: We evaluate intelligibility using Word Error Rate (WER) on the Seed-TTS\cite{anastassiou2024seed} test set. Acoustic fidelity and similarity are measured via Speaker Similarity\footnote{\url{https://github.com/resemble-ai/Resemblyzer}}, Emotion Similarity\footnote{\url{https://huggingface.co/emotion2vec}}, Log-Spectral Distance (LSD), Mel-Cepstral Distortion (MCD), Mean Squared Error of Pitch (MSEP), and Voiced/Unvoiced Mismatch Rate (MR). Music Metrics: We utilize the SongEval\cite{yao2025songeval} benchmark to assess musical attributes including coherence, musicality, and memorability. Sound Effect Metrics: We adopt standard TTA metrics on the AudioCaps test set. Fréchet Audio Distance (FAD) for audio quality, Fréchet Distance (FD) based on PANNs\cite{kong2020panns}, Inception Score (IS) for generation diversity, and CLAP Score\cite{wu2023large} for text-audio alignment. Control \& Subjective Metrics: We assess control capability via Classification Control Accuracy through human listening tests, where annotators verify if generated samples match specific attributes (e.g., Age, Genre, Atmosphere). Subjective quality is evaluated using Quality Mean Opinion Score (QMOS), Naturalness MOS (NMOS), and Musicality MOS (MMOS).

\subsection{Subjective Evaluation Details}
\label{sec:appendix:subjective_details}

To rigorously assess the perceptual quality of the synthesized audio, we conducted subjective listening tests following the standard Mean Opinion Score (MOS) protocol.

We recruited 20 volunteer listeners with normal hearing. For speech evaluation, all participants were native speakers of Chinese. To ensure consistent acoustic conditions, participants were provided with high-quality monitoring headphones and instructed to perform the evaluation in a quiet, sound-isolated environment.

Participants rated samples on a 5-point Likert scale (1 = Bad, 5 = Excellent, with 0.5 increments). The evaluation focused on three distinct dimensions corresponding to the unified tasks:
Naturalness MOS (NMOS):Evaluated specifically for speech (TTS), focusing on prosody, intonation, and human-like articulation.
Musicality MOS (MMOS):Evaluated for music (TTM), focusing on melodic coherence, rhythmic stability, and harmony.
Quality MOS (QMOS):Evaluated across all modalities (including SFX), focusing on

\subsection{Test Sets}
For speech, we use the complete Seed-TTS test set for WER and a manually annotated set of 500 instruction-phoneme pairs for control evaluation. For music, we construct a 500-sample test set with descriptions covering genre, instrument, and atmosphere. For sound effects, evaluations are conducted on the standard AudioCaps test split.

\end{document}